\documentclass[10pt,letterpaper]{article}

\usepackage{opex3}
\usepackage{graphics}
\usepackage{graphicx}

\begin{document}

\title{Intense ultra-broadband down-conversion from randomly poled nonlinear crystals}

\author{Ji\v{r}\'{\i} Svozil\'{\i}k and  Jan Pe\v{r}ina Jr.}
\address{Joint Laboratory of Optics, Palack\'{y} University
and Institute of Physics of Academy of Science of the Czech
Republic, 17. listopadu 50A, 772 07 Olomouc, Czech Republic}
\email{perinaj@prfnw.upol.cz}

\begin{abstract}
Randomly poled nonlinear crystals are shown to be able to emit
intense ultra-broadband photon-pair fields with properties
comparable to those coming from chirped periodically-poled
crystals. Their intensities scale linearly with the number of
domains. Also photon pairs extending over intervals with durations
comparable to one optical cycle can be generated in these
crystals.
\end{abstract}

\bibliographystyle{osajnl}

\section{Introduction}

Spontaneous parametric down-conversion (SPDC) with its production
of photon pairs belongs to the most fascinating nonlinear optical
effects. Two photons comprising a photon pair can be entangled in
their degrees of freedom as it was first observed by Hong, Ou, and
Mandel \cite{Hong1985} for temporal correlations. Nonlinear bulk
crystals served nearly exclusively as sources of these photon
pairs for many years. However, many highly nonlinear crystals
could not be used due to impossibility to achieve natural
phase-matching conditions. Here, the concept of additional
periodic modulation of nonlinear susceptibility as introduced by
Armstrong \cite{Armstrong1962} has become fruitful and resulted in
the invention of poling of nonlinear crystals \cite{Hum2007}.
Highly nonlinear materials in which
quasi-phase-matching conditions \cite{Fejer1992} are met can be
efficiently used since then. They have allowed the construction of
bright and versatile sources of photon pairs.

Moreover, periodical poling has also allowed to tailor the properties of emitted photon pairs
using nonlinear domains with variable lengths (chirped periodical poling). Domains of
different lengths in an ordered structure allow an efficient nonlinear interaction in an
ultra-wide spectral region extending typically over several hundreds of nm
\cite{Nasr2008,Harris2007,Svozilik2009,Saleh2009}. It has been shown that such photon pairs
can posses quantum temporal correlations at the timescale of fs and so can be extraordinarily
useful, e.g., in metrology (quantum optical coherence tomography \cite{Carrasco2004}) or
quantum-information processing \cite{Humble2007}.

Alternatively, domains of different lengths can be ordered randomly. The nonlinear interaction
can be efficient even in this case [stochastic quasi-phase-matching] as studies of the process
of second-harmonic generation indicate
\cite{Fejer1992,Morozov2004,Vidal2006,Aleksandrovsky2008,Kitaeva2007}. Here, we show
that these structures despite their randomness are able to generate spectrally ultra-wide
photon pairs at generation rates comparable to chirped periodically-poled crystals (CPPC).
This shows that contributions from the coherent summation of photon-pair fields coming from
ordered domains of different lengths in CPPCs are considerably lower than generally
accepted. When properly phase compensated photon pairs with correlation times at the fs
timescale can be emitted from these structures. Moreover, these structures have usually
smaller requirements with respect to fields' polarization properties and orientation of the
nonlinear medium \cite{Baudrier-Raybaut2004}. Also fabrication tolerances are less strict in
this case \cite{Fejer1992}. Practically, technology of poling produces structures with
relatively large declinations from an ideal geometry which results in reduction of photon-pair
generation rates. There is the hope that the fabricated randomly poled structures can even
out-perform their chirped periodically-poled counterparts. The extension to 2D or 3D
geometries offers additional possibilities including naturally poled materials
\cite{Baudrier-Raybaut2004}.

\section{Spontaneous parametric down-conversion in poled nonlinear crystals}
Quantum state $ |\psi\rangle $ describing a photon pair generated in the process of SPDC in a
poled crystal can be written as follows \cite{Hong1985}:
\begin{equation}   
 |\psi\rangle = \int d\omega_s \int d\omega_i \Phi(\omega_s,\omega_i)
  \hat{a}_s^\dagger(\omega_s) \hat{a}_i^\dagger(\omega_i)
  |{\rm vac} \rangle ,
\label{1}
\end{equation}
where a two-photon spectral amplitude $ \Phi $ gives the probability amplitude of emitting a
signal photon at frequency $ \omega_s $ and its idler twin at frequency $ \omega_i $. It can
be obtained in the form:
\begin{equation}   
 \Phi(\omega_s,\omega_i) = g(\omega_s,\omega_i)
  \xi_p F(\Delta k(\omega_s,\omega_i)) \delta(\omega_p^0-\omega_s-\omega_i)
\label{2}
\end{equation}
assuming cw pumping with amplitude $ \xi_p $, frequency $
\omega_p^0 $ and wave-vector $ k_p $. Creation operator $
\hat{a}_a^\dagger(\omega_a) $ in Eq.~(\ref{1}) generates a photon
with wave-vector $ k_a $ and frequency $ \omega_a $ ($ a=s,i $)
into the vacuum state $ |{\rm vac} \rangle $. Coupling constant $
g $ is given as $ g(\omega_s,\omega_i) = \chi^{(2)}(0)
\sqrt{\omega_s\omega_i} / [ic\pi\sqrt{
n_s(\omega_s)n_i(\omega_i)}] $ where $ \chi^{(2)} $ denotes
second-order susceptibility, $ n_a(\omega_a) $ means index of
refraction, and $ c $ is speed of light in vacuum. Symbol $ \Delta
k $ stands for nonlinear phase mismatch ($ \Delta k = k_p - k_s -
k_i $). The stochastic phase-matching function $ F $ introduced in
Eq.~(\ref{2}) takes the following form:
\begin{eqnarray}   
 F(\Delta k) = \sum_{n=1}^{N_L} (-1)^{n-1}\int_{z_{n-1}}^{z_{n}} dz
  \exp(i\Delta k z) .
\label{3}
\end{eqnarray}
In Eq.~(\ref{3}), symbol $ N_L $ gives the number of domains and $
n $-th domain extends from $ z=z_{n-1} $ to $ z=z_n $. Positions $
z_n $ of domain boundaries are random and can be described as $
z_n = z_{n-1} + l_0 + \delta l_n $ ($ n=1,\ldots,N_L $, $ z_0 = 0
$) using stochastic declinations $ \delta l_n $. The basic domain
length $ l_0 $ is determined such that quasi-phase-matching is
reached, i.e. $ l_0 = \pi/ \Delta k_0 $, $ \Delta k_0 \equiv
\Delta k(\omega_s^0,\omega_i^0) $ and $ \omega_a^0 $ stands for
the central frequency of field $ a $. Independent random
declinations $ \delta l_n $ are assumed to obey the joint Gaussian
probability distribution $ P $:
\begin{equation}    
 P(\delta {\bf L})=\frac{1}{ (\sqrt{\pi} \sigma)^{N_L} }
  \exp(-\delta {\bf L}^T {\bf B} \delta {\bf L} ).
\label{e}
\end{equation}
Covariance matrix $ {\bf B} $ is diagonal and its nonzero elements
are equal to $ 1 / \sigma^2 $. Stochastic vector $ \delta {\bf L}
$ is composed of declinations $ \delta l_n $; symbol $ {}^T $
stands for transposition.

Integration of the expression in Eq.~(\ref{3}) leaves us with the
following simple formula (valid for $ N_L \gg 1 $):
\begin{equation}   
 F(\Delta k) = \frac{2i}{\Delta k} \sum_{j=0}^{N_L} (-1)^j \exp(i\Delta
  k z_j) .
\label{5}
\end{equation}
This formula can be interpreted such that SPDC occurs only in
domains with positive susceptibility $ \chi^{(2)} $ with doubled
amplitudes and domains with negative susceptibility $ \chi^{(2)} $
play only the role of a 'linear' filler. This elucidates why two
neighbouring domains form basic elementary units for the
interpretation of properties of photon-pairs \cite{Svozilik2009}.

The emitted photon pairs can be characterized by mean spectral
density $ n(\omega_{s},\omega_{i}) $ of the number of photon
pairs. The density $ n $ is defined along the formula $
n(\omega_{s},\omega_{i})=\langle \langle \psi|
\hat{a}^{\dagger}_s(\omega_s)\hat{a}_s(\omega_s)\hat{a}^\dagger_i(\omega_i)
\hat{a}_i(\omega_i) |\psi \rangle \rangle_{\rm av} $ where the
symbol $ \langle \rangle_{\rm av} $ means stochastic averaging
over an ensemble of random realizations of the crystal. In
practice mean values can correspond to averaging over states of
photon pairs emitted at different positions in the transverse
plane of a naturally-poled material \cite{Baudrier-Raybaut2004}.
Or they can just serve as an indicator of expected values for
individual realizations. The spectral density $ n $ in quantum
state $ |\psi\rangle $ reads:
\begin{eqnarray}    
 n(\omega_{s},\omega_{i}) &=& \frac{|g(\omega_{s},\omega_{i})|^{2} |\xi_{p}|^{2}
}{2\pi}
  \langle |F(\Delta k(\omega_{s},\omega_i))|^{2}\rangle_{\rm
av}\delta(\omega_p^0-\omega_s-\omega_i).
\label{6}
\end{eqnarray}

In the considered random structures, the averaged squared modulus
of phase-matching function $ F $, defined as $\int d\delta {\bf L}P(\delta {\bf
L})|F(\Delta k)|^2 $, is obtained as follows:
\begin{eqnarray}   
 \langle |F(\Delta k)|^2 \rangle_{\rm av} &=& \frac{4 }{
  \Delta k^2 } \left( (N_L+1) \frac{1-|H(\delta k)|^2}{|1-H(\delta k)|^2 } -
  \left[ \frac{ H(\delta k)[1-H(\delta k)^{N_L+1}] } {[1-H(\delta k)]^2} + {\rm c.c.}
  \right]\right) ;
\label{7}
\end{eqnarray}
$ \delta k(\omega_s,\omega_i) = \Delta k(\omega_s,\omega_i) -
\Delta k_0 $. Symbol $ {\rm c.c.} $ replaces the
complex-conjugated term and $ H(\delta k) = \exp[ i\delta k l_0 ]
\exp[ -(\sigma \Delta k)^2/4 ] $. A detailed analysis of the
formula in Eq.~(\ref{7}) shows that $ \langle |F(\Delta k)|^2
\rangle_{\rm av} $ is peaked around $ \delta k = 0 $. The larger
the deviation $ \sigma $ the broader the peak.

We consider CPPC for comparison. In this case, the ordered
positions of boundaries are given as $ z_n = n l_0 + \zeta'
(n-N_L/2)^2 l_0^2 $, $ \zeta' = \zeta/ \Delta k_0 $, and $ \zeta $
stands for a chirping parameter. The phase-matching function $
F^{\rm chirp} $ can then be derived in the form \cite{Harris2007}:
\begin{equation}   
 F^{\rm chirp}(\Delta k) = \frac{2\sqrt{\pi} }{
  \sqrt{i\zeta' \Delta k^3 l_0 } }\exp\left(\frac{i\Delta k N_L l_0}{2}\right)\exp\left(
  -\frac{i\delta k^2}{4\Delta k \zeta'} \right) \left|
  {\rm erf}(f(N_L/2)) - {\rm erf}(f(-N_L/2))
  \right|^2,
\label{8}
\end{equation}
$ f(x) = (\sqrt{-i}/2) (\sqrt{\zeta' \Delta k} x l_0 + \delta k/
\sqrt{\zeta' \Delta k} ) $ and $ {\rm erf} $ means the error
function. It holds that the larger the chirping parameter $ \zeta
$ the broader the phase-matching function $ F^{\rm chirp}(\Delta
k) $.

\section{Photon-pair generation rates and intensity spectra}
In our investigation, we use spectrally degenerate collinear
down-conversion from a poled LiNbO$ {}_3 $ crystal pumped at the
wavelength $\lambda_p^0 =775 $~nm. The signal and idler photons
occur at the fiber-optics communication wavelength $ \lambda_s =
\lambda_i = 1.55~\mu $m. The crystal optical axis is perpendicular
to the fields' propagation direction and is parallel to the
vertical direction. All fields are vertically polarized and so the
largest nonlinear element $ \chi^{(2)}_{33} $ is exploited. The
basic domain length $ l_0 $ equals to 9.515~$ \mu $m. A structure
composed of $ N_L = 2000 $ layers is roughly 19~mm long and
typically delivers $ 2 \times 10^7 $ photon pairs per second per
100~mW of pumping for ordered positions of boundaries.

\begin{figure}   
\centering
\includegraphics[scale=0.24]{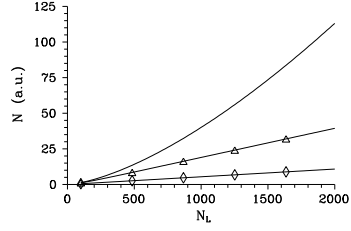}

 \caption{Photon-pair generation rate $ N $ as a function of the number $ N_L $ of
 domains for an ensemble of random crystals with standard deviation $ \sigma $ equal
 to 0~m (solid line),
 0.5~$ \times10^{-6}$~m (solid line with $ \triangle $), and 2~$ \times10^{-6}$~m (solid line
 with $ \diamond $);
 $  N = \int d\omega_s\int d\omega_i
 n(\omega_s,\omega_i) $.}
\label{fig1}
\end{figure}

\begin{figure}   
\centering
a)
 \includegraphics[scale=0.19]{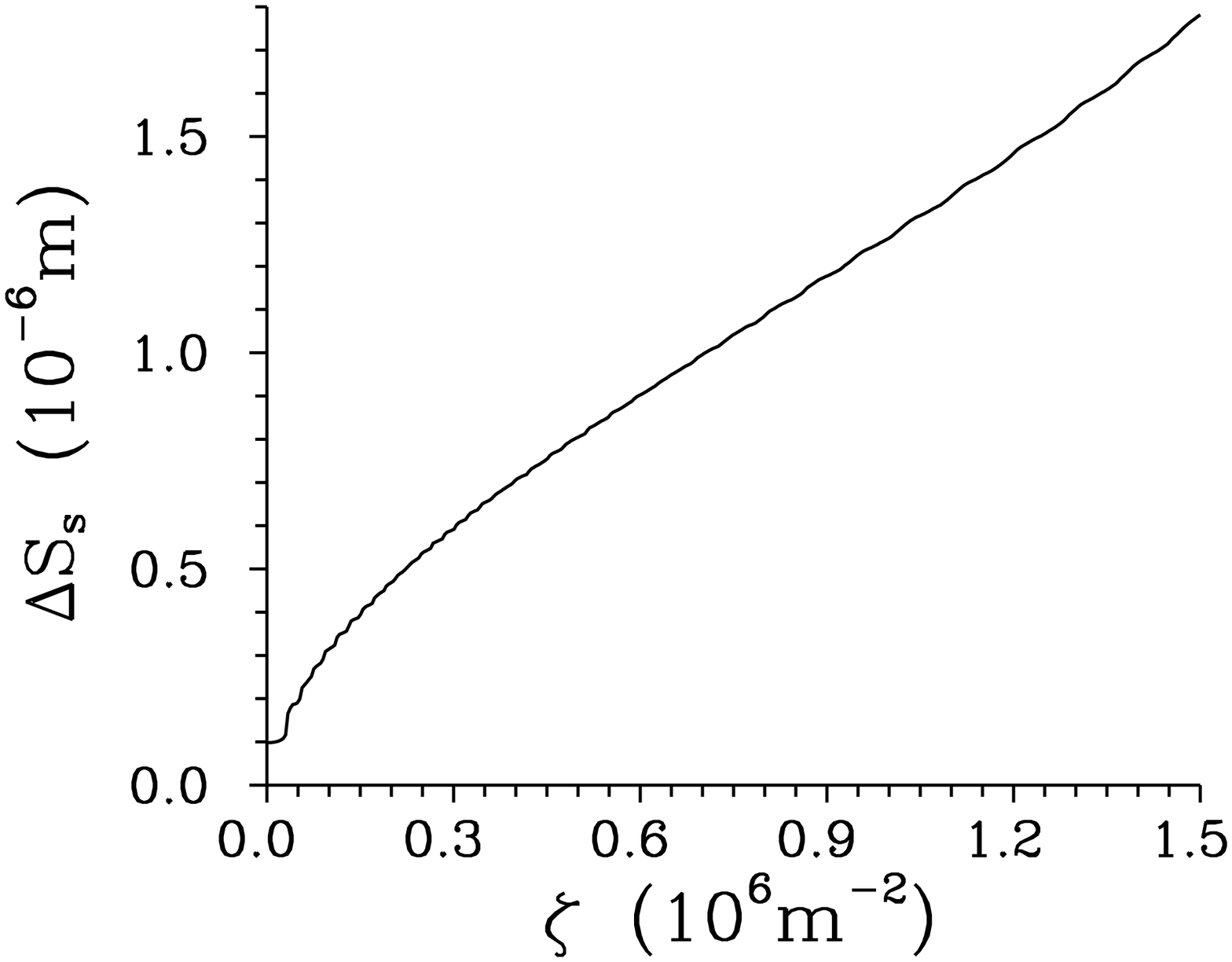}
b)
 \includegraphics[scale=0.19]{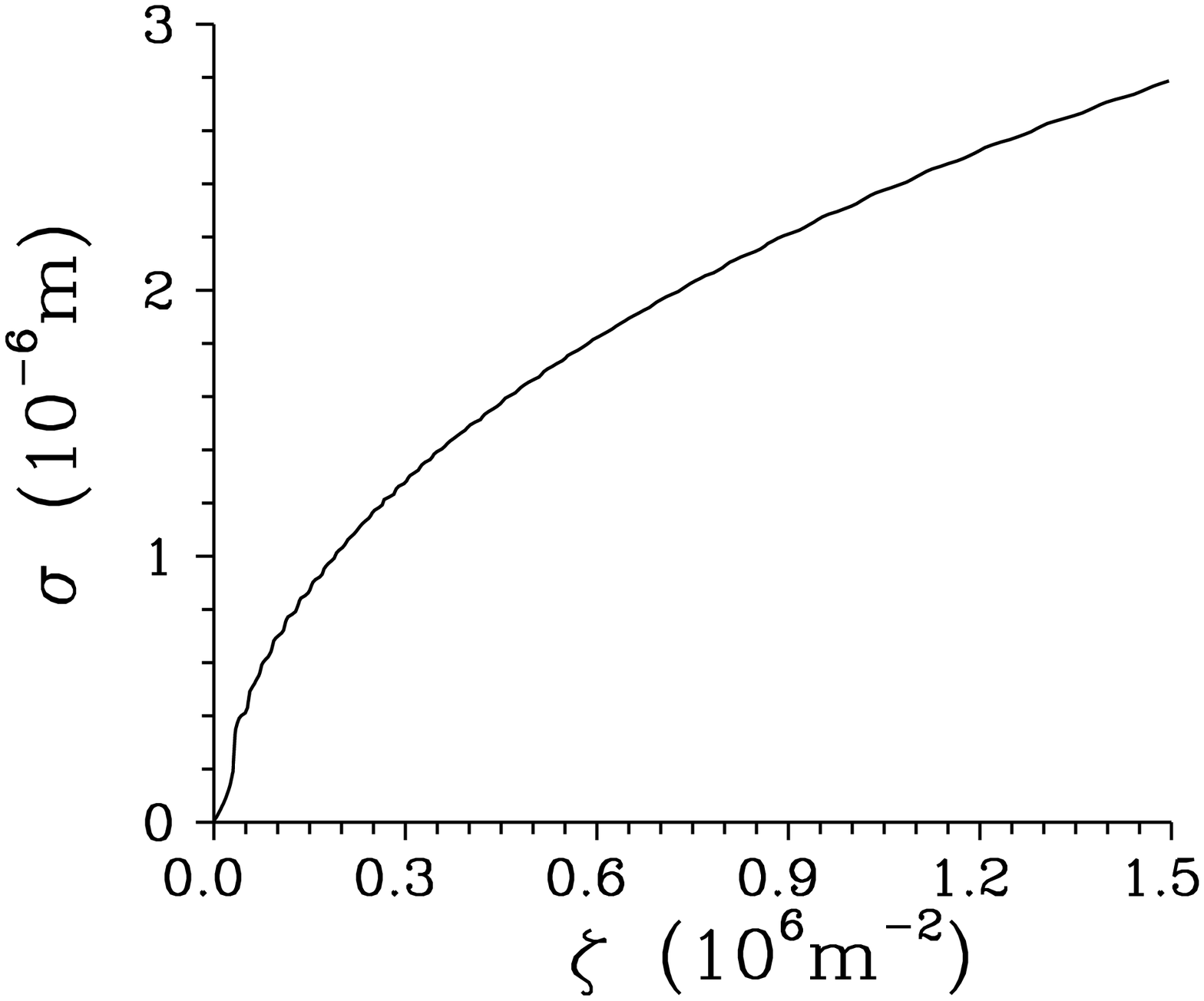}
c)
\includegraphics[scale=0.23]{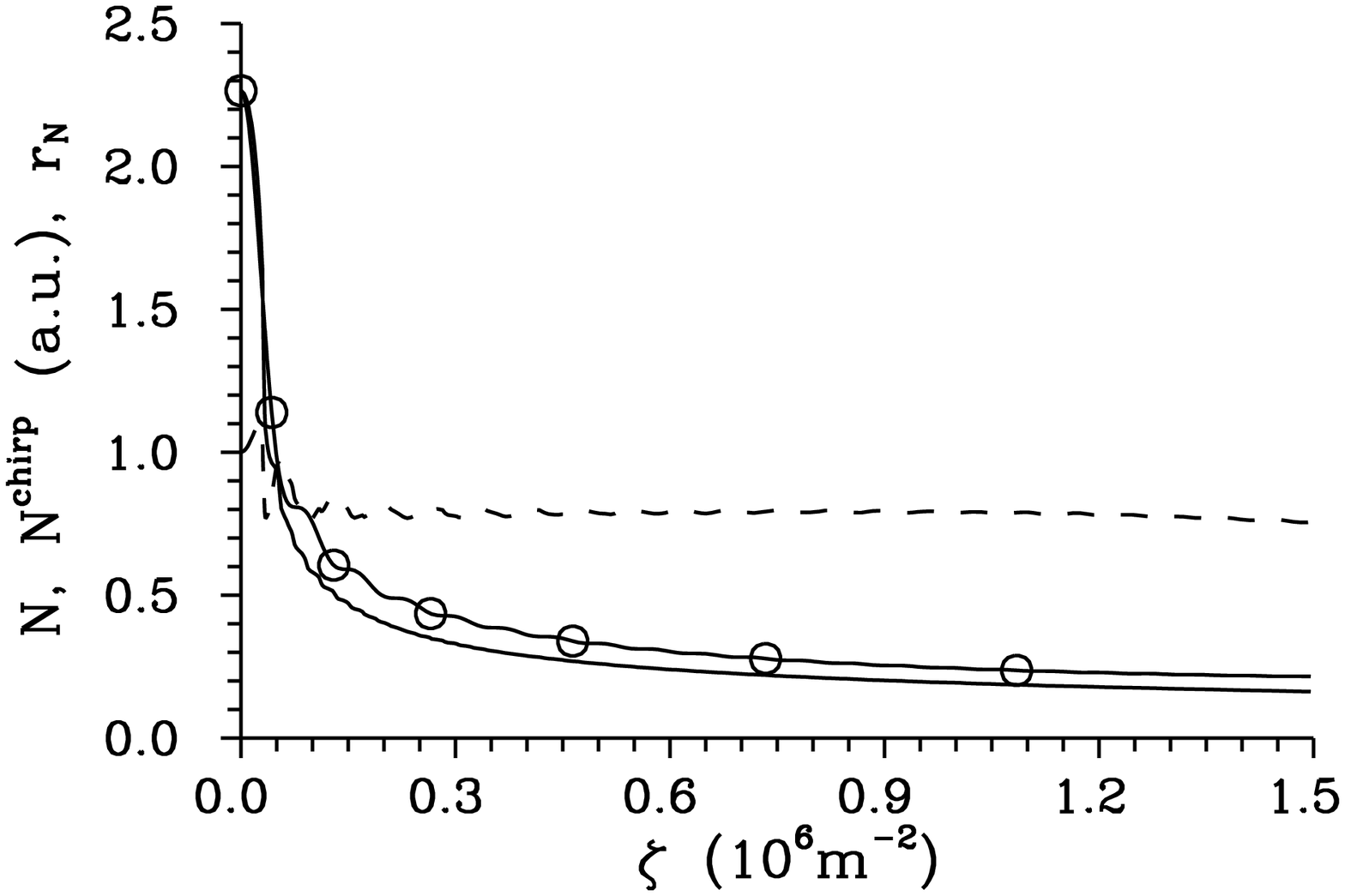}
 \caption{a) Signal-field spectral width $ \Delta S_s $ (FWHM) as a function of
  chirping parameter $ \zeta $, b) transformation curve between the
  standard deviation $ \sigma $ and chirping
  parameter $ \zeta $ assuming equal spectral widths $ \Delta S_s $, and c)
  photon-pair generation rate $ N $ for chirped (solid curve  with $ \circ $) and
  random (solid curve) crystals and their ratio $
  r_N $ ($ r_N = N / N^{\rm chirp} $, dashed curve) as functions
  of chirping parameter $ \zeta $; $ N_L=2000 $.}
\label{fig2}
\end{figure}

The most striking feature of random structures is that the
photon-pair generation rate $ N $ increases linearly with the
number $ N_L $ of domains (see Fig.~\ref{fig1}). On the other
hand, standard deviation $ \sigma $ plays the central role in the
determination of spectral widths $ \Delta S_s $ and $ \Delta S_i $
of the signal and idler fields. The larger the deviation $ \sigma
$ the broader the signal- and idler-field spectra $ S_s $ and $
S_i $ (see Fig.~\ref{fig2}a and Fig.~\ref{2}b for $ \sigma - \zeta
$ transformation). This can be understood from the behavior of
spatial spectrum of the $ \chi^{(2)}(z) $ modulation participating
in phase matching conditions: the larger the deviation $ \sigma $
the broader the spatial spectrum of $ \chi^{(2)}(z) $. However,
the spectral broadening is at the expense of photon-pair
generation rates $ N $ (see Fig.~\ref{fig2}c).

Similar behavior as observed in random structures has been found
in CPPCs \cite{Harris2007,Svozilik2009} considering the chirping
parameter $\zeta$ instead of the deviation $\sigma$. Our
investigations have revealed that this similarity is both
qualitative and quantitative (see Fig.~\ref{2}). For any value of
the chirping parameter $ \zeta $ there exists a value of the
standard deviation $ \sigma $ such that the spectral widths $
\Delta S_s $ and $ \Delta S_i $ equal. Moreover, also the
photon-pair generation rates $ N $ are comparable. This behavior
is illustrated in Fig.~\ref{fig2} for structures with $ N_L=2000 $
domains. The signal-field spectra $ S_s $ in CPPCs are
extraordinarily wide (larger that 1~$\mu $m) for sufficiently
large values of parameter $ \zeta $ (see Fig.~\ref{fig2}a). The
signal-field spectra $ S_s $ coming from random structures can
have the same widths assuming sufficiently strong randomness: the
transformation curve between the standard deviation $ \sigma $ and
chirping parameter $ \zeta $ conditioned by equal spectral widths
$ \Delta S_s $ is plotted in Fig.~\ref{fig2}b. The photon-pair
generation rates $ N $ for random and chirped crystals are
compared in Fig.~\ref{fig2}c. Photon-pair generation rates $ N $
in random crystals reach at least 80~\% of those of CPPCs.

\begin{figure}[h]   
\centering
a)
 \includegraphics[scale=0.19]{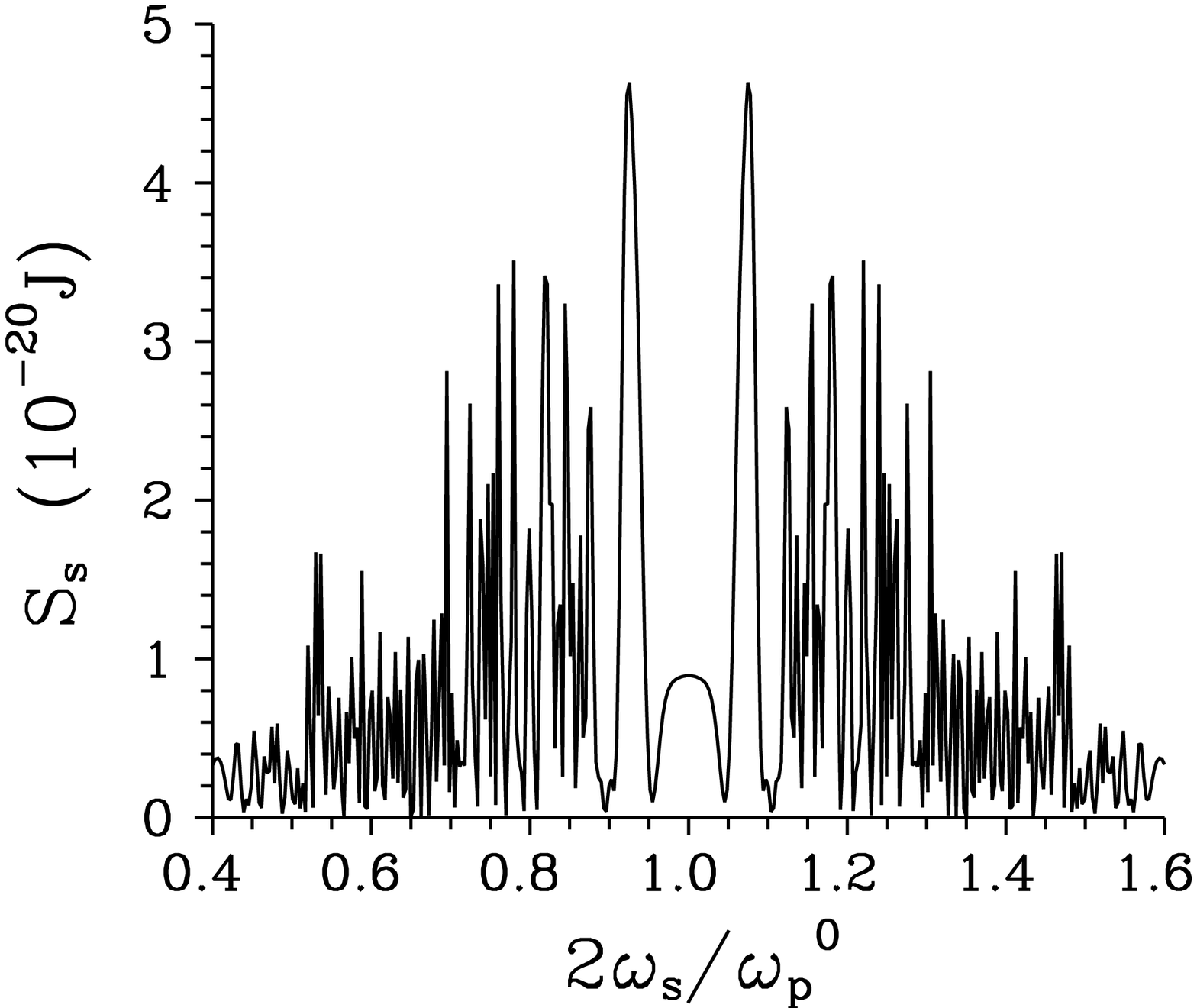}
\hspace{1cm}
b)
 \includegraphics[scale=0.19]{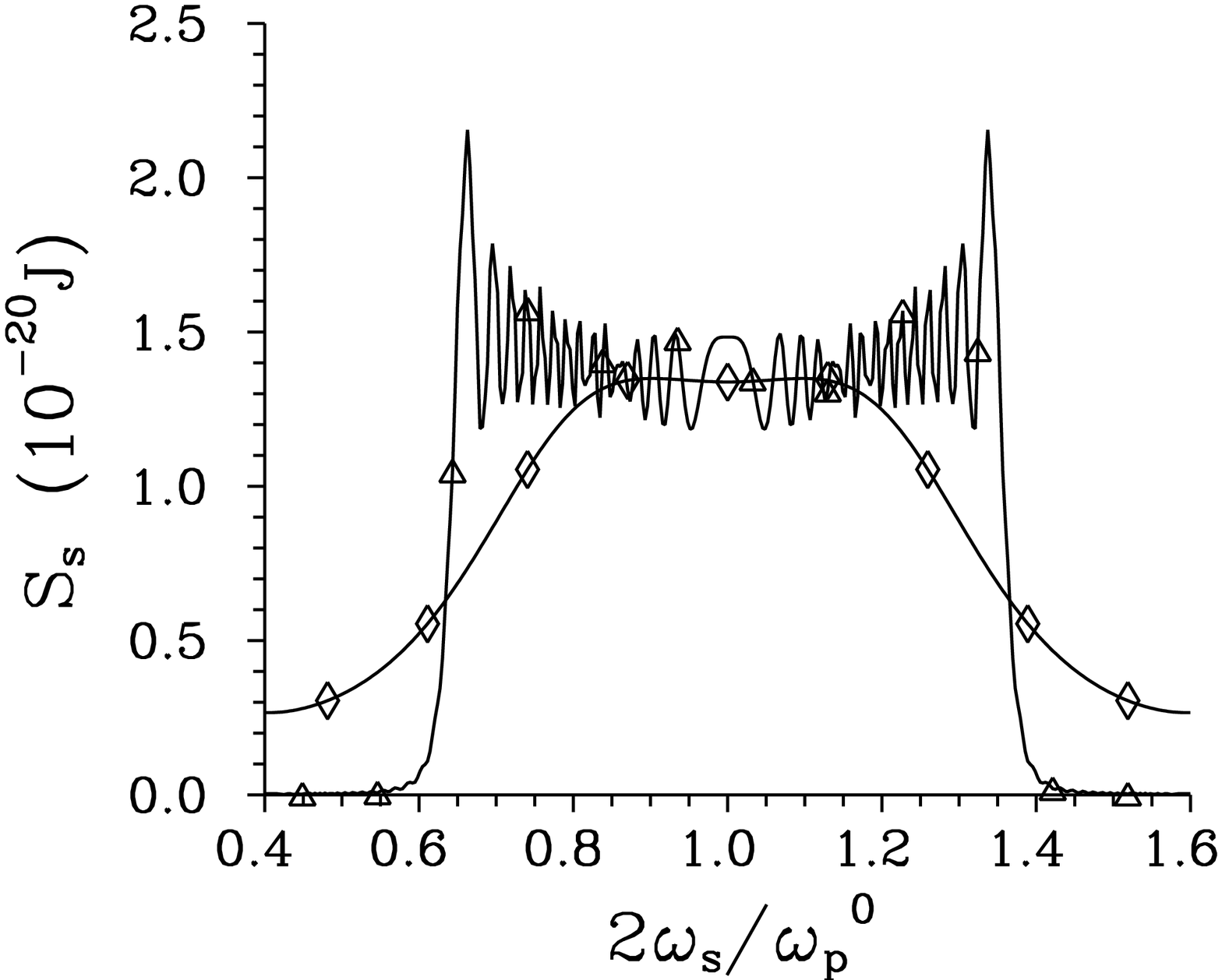}
 \caption{Signal-field spectrum $ S_s $ for
  a) one realization of the random crystal and b)
  CPPC (solid curve with $ \triangle $) and mean value for an ensemble of random
  crystals (solid curve with $ \diamond $). Spectra $ S_s $ are given as
  $ S_{s} = \hbar\omega_s\int d\omega_i n(\omega_s,\omega_i) $ and are normalized
  such that one photon is emitted;
  $ \sigma = 2.3 $~$ \times 10^{-6} $~m, $ \zeta = 1$~$\times 10^{6}
  $~m$^{-2} $, $ N_L=2000 $.}
\label{fig3}
\end{figure}

Comparison of the signal-field spectra $ S_s $ for one typical
realization of the random structure and a CPPC (see
Fig.~\ref{fig3}) reveals that spectra of random structures cover a
larger range of frequencies and are composed of many local peaks.
Nevertheless, the averaged spectrum of the random crystal has the
same FWHM as the considered CPPC (see Fig.~\ref{3}b).

\section{Temporal correlations}

\begin{figure}   
\centering
a)
 \includegraphics[scale=0.19]{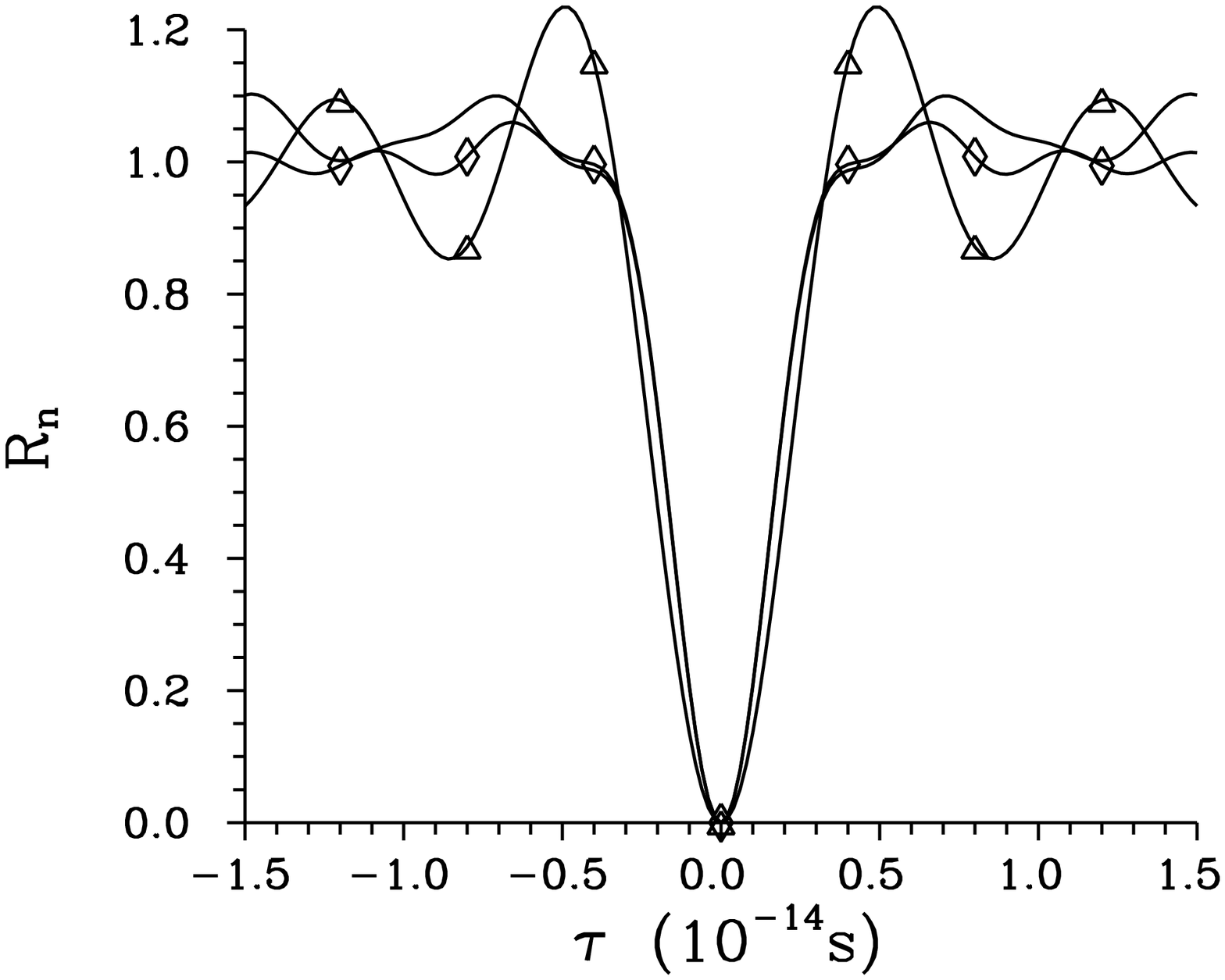}
\hspace{1cm}
b)
 \includegraphics[scale=0.19]{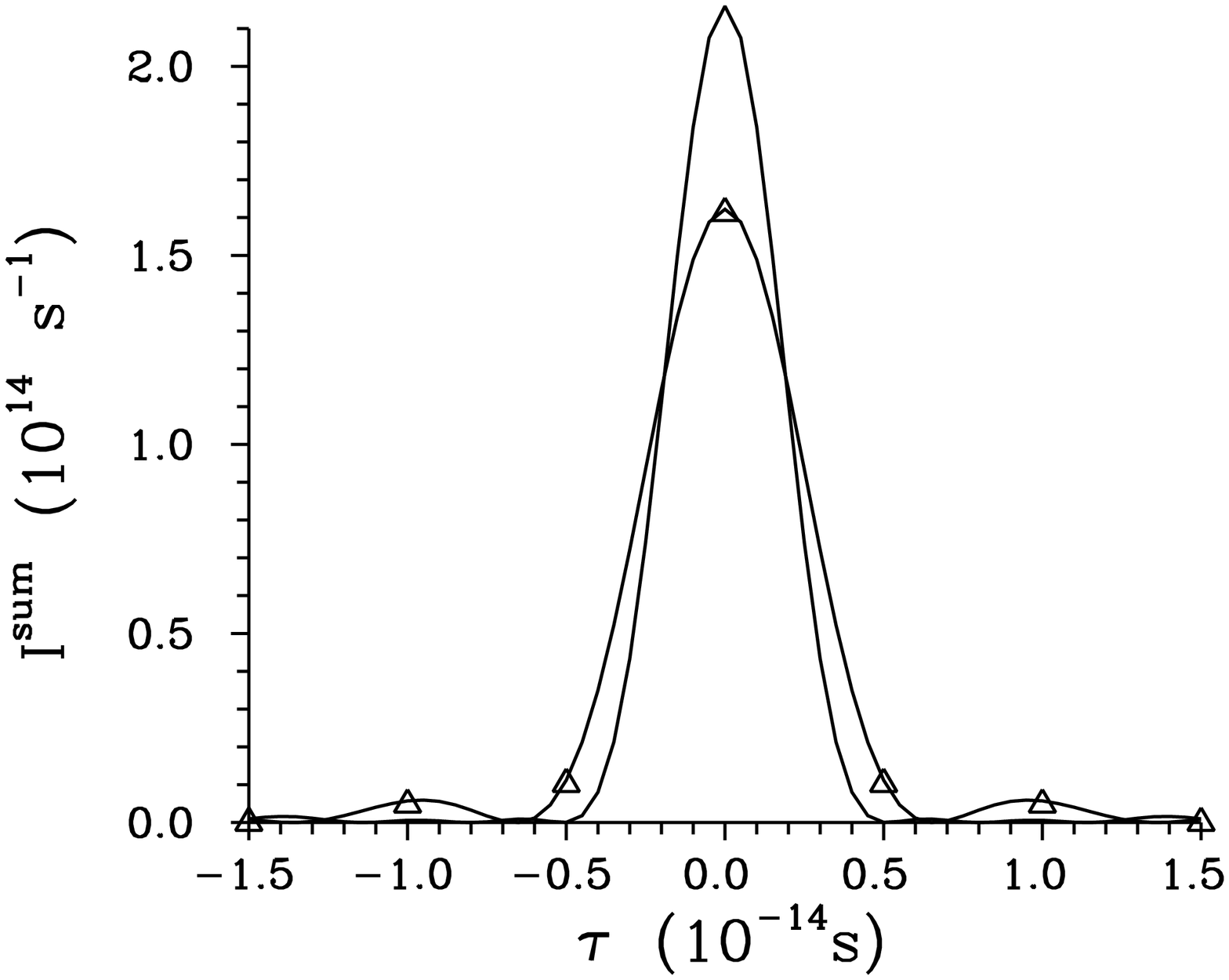}
 \caption{a) Coincidence-count rate $ R_n $ and b) sum-frequency field intensity $ I^{\rm sum} $
  as they depend on relative time delay $ \tau $ for one realization of the random crystal (solid curve),
  chirped crystal (solid curve with $ \triangle $)
  and an ensemble of random crystals (solid curve with $ \diamond
  $). In b), ideal phase compensation is assumed and curves are normalized
  such that $ \int_{-\infty}^{\infty}  d\tau I^{\rm sum}(\tau) = 1 $;
  values of parameters are the same as in Fig.~\ref{fig3}. }
\label{fig4}
\end{figure}

These ultra-wide spectra allow to generate photon pairs with
extremely short temporal correlations that can be measured in a
Hong-Ou-Mandel interferometer \cite{Hong1985}. The
coincidence-count rate $ R_n $ in this interferometer as a
function of relative time delay $ \tau $ between two photons forms
a typical dip (see Fig.~\ref{fig4}) that can be described by the
formula:
\begin{eqnarray}   
 & R_n(\tau) = 1 - \frac{1}{R_0}{\rm Re} \Biggl[ \exp(i\omega_p^0\tau) \int
  d\omega_s\exp(-2i\omega_s\tau) \langle \left| F\left(\Delta
  k(\omega_s,\omega_p^0-\omega_s)
   \right) \right|^2 \rangle_{\rm av} \Biggr] ,
\label{9} \\
 & R_0 =  \int d\omega_s
  \langle \left| F\left(\Delta k(\omega_s,\omega_p^0-
  \omega_s)\right) \right|^2 \rangle_{\rm av}.
\label{10}
\end{eqnarray}
Entanglement times derived from the width of the coincidence-count
dip can be as short as several fs for both random and chirped
structures. Each realization of the random crystal leads to a
sharp peak as evident from the profile of the averaged
coincidence-count rate $ R_n $ in Fig.~\ref{fig4}a. They mainly
differ in amplitudes of oscillations occurring at the shoulders.

The behavior observed in the interferometer shows the potential to
generate photon pairs with wave-packets extending over the period
of only several fs provided that a good phase compensation is
reached. The temporal profile of a wave-packet is given by the
fourth-order stationary correlation function $ I_{\Phi}(\tau) $
defined as
\begin{equation}    
 I_{\Phi}(\tau) = \frac{1}{2T} \int_{-T}^{T}dt \left| \langle
  {\rm vac} | \hat{E}_s^{(+)}(t) \hat{E}_i^{(+)}(t-\tau)
  |\psi\rangle \right|^2 ,
\label{11}
\end{equation}
where $ T $ denotes detection duration. Function $ I_{\Phi} $
gives the probability of detecting an idler photon in the instant
that precedes the instant of signal-photon detection by $ \tau $.
In experiment, function $ I_{\Phi}(\tau) $ is obtained from the
measurement of intensity $ I^{\rm sum} $ of the sum-frequency
field generated by two photons mutually delayed by $ \tau $ [$
I_{\Phi}(\tau) \propto I^{\rm sum}(\tau)$]. Intensities $ I^{\rm
sum} $ created by photon pairs coming from the considered ideally
phase-compensated structures are plotted in Fig.~\ref{fig4}b and
demonstrate the principal ability to detect both photons in a
window $ \approx 5 $~fs wide. Restricting ourselves to quadratic
phase compensation \cite{Harris2007,Brida2009,Sensarn2010}, this
window is roughly two times wider. Quadratic compensation is more
powerful in CPPCs, however, the difference is not large. This
emphasizes the potential of the studied photon pairs as a tool of
diagnostics of ultra-fast processes in physics, biology or
chemistry.

\section{Conclusion}

Quantitative similarity in properties of photon pairs generated
from randomly poled and chirped periodically-poled crystals has
been found. Namely ultra-wide signal and idler fields can be
emitted from randomly poled crystals. Two photons in a pair can in
principle occur together in a temporal window that characterizes
one optical cycle. Importantly, photon-pair generation rates are
comparable and depend linearly on the number of domains. Contrary
to chirped periodically-poled crystals the randomly poled crystals
are quite tolerant in fabrication. This gives a great promise for
the use of randomly poled crystals as standard bright ultra-wide
spectral sources of entangled photon pairs.

\section*{Acknowledgement}
Support by projects IAA100100713 of GA AV\v{C}R, COST OC 09026,
1M06002 of M\v{S}MT and PrF-2010-009 of Palacky University is
acknowledged. The authors thank O. Haderka for discussions.
\end{document}